\documentclass[runningheads,dvipsnames]{llncs}
\usepackage[english]{babel}
\usepackage{hyperref}
\usepackage{multirow} %
\usepackage{subcaption} %
\usepackage{caption}
\usepackage{subcaption}
\usepackage{microtype}
\usepackage{booktabs} %
\usepackage[table,xcdraw]{xcolor} %
\usepackage{graphicx} %
\usepackage{mathtools} %
\usepackage{amssymb} %
\usepackage{tabularx}
\usepackage{paralist}
\usepackage{enumitem}
\usepackage[T1]{fontenc}
\usepackage{wrapfig} %
\usepackage{tabularray}
\usepackage{changepage}

\usepackage{tikz, calc, tikz-qtree}
\usepackage{xfrac} %

\newcommand{\processTreeOp}{\ensuremath{\oplus}} %

\newcommand{\mycomment}[1]{} %

\newcommand{\mypar}[1]{\smallskip\noindent\textbf{#1.}}

\DeclarePairedDelimiter\abs{\lvert}{\rvert} %

\hypersetup{
	pdftoolbar=true,        %
	pdfmenubar=true,        %
	pdffitwindow=false,     %
	pdfstartview={FitH},    %
	pdftitle={},    %
	pdfauthor={},     %
	pdfsubject={},   %
	pdfcreator={},   %
	pdfproducer={}, %
	pdfkeywords={}, %
	pdfnewwindow=true,      %
	colorlinks=true,       %
	  linkcolor=Brown,          %
	  citecolor=OliveGreen,        %
	  filecolor=magenta,      %
	  urlcolor=NavyBlue           %
}

\begin{document}
\renewcommand*\sectionautorefname{Section}
\renewcommand*\subsectionautorefname{Section}
\newcommand*\exampleautorefname{Example}
\newcommand*\lemmaautorefname{Lemma}
\newcommand*\propertyautorefname{Property}
\newcommand*\problemautorefname{Problem}
\newcommand*\definitionautorefname{Def.}
\renewcommand\equationautorefname{Eq.}
\renewcommand*\figureautorefname{Fig.}

\title{Control-flow Reconstruction Attacks on Business Process Models}

\titlerunning{Control-flow Reconstruction Attacks on Business Process Models}

\author{Henrik Kirchmann\inst{1} \and Stephan A. Fahrenkrog-Petersen\inst{1,2} \and Felix Mannhardt \inst{3} \and Matthias Weidlich \inst{1}}

\authorrunning{H. Kirchmann et al.}

\institute{
Humboldt-Universit\"at zu Berlin, Berlin, Germany\\
\email{\{henrik.kirchmann, stephan.fahrenkrog-petersen, matthias.weidlich\}@hu-berlin.de}
\and
Weizenbaum Institute for the Networked Society, Berlin, Germany
\and
Eindhoven University of Technology, Eindhoven, Netherlands\\
\email{f.mannhardt@tue.nl}\\
}

\maketitle

\begin{abstract}

Process models may be automatically generated from event logs that
contain as-is data of a business process. While such models generalize
over the control-flow of specific, recorded process executions, they are
often also annotated with behavioural statistics, such as execution
frequencies.
Based thereon, once a model is published, certain insights about the
original process executions may be reconstructed, so that an external party
may extract confidential information about the business process.
This work is the first to empirically investigate such reconstruction
attempts based on process models. To this end, we propose different play-out
strategies that reconstruct the control-flow from process trees, potentially
exploiting frequency annotations. To assess the potential success
of such reconstruction attacks on process models, and hence the risks
imposed by publishing them, we compare the reconstructed process executions
with those of the original log for several real-world datasets.

\keywords{Reconstruction Attacks\and Process Analysis \and Model Play-out}
\end{abstract}

\section{Introduction}
Under the umbrella of process mining, event logs that have been recorded by information systems facilitate the analysis
of qualitative and quantitative properties of business processes~\cite{vanDerAalst2016process}.
Event logs support information systems engineering through the discovery of
process
models~\cite{DBLP:journals/tkde/AugustoCDRMMMS19},
which are useful for understanding the flow of the process and, once annotated with performance characteristics, help to identify performance bottlenecks and improvement opportunities.

Discovery algorithms generalize and aggregate the behaviour recorded in an
event log. As a consequence, individual process executions are not directly
represented, when publishing the model~\cite{maatouk2022quantifying}, e.g.,
to an external party for the
purpose of process certification, staff training or consulting. However, in practice,
process models are not limited to the
generalized control-flow of a process. Rather, they also contain
summary statistics about the behaviour, most prominently execution
frequencies or branching
probabilities~\cite{burke2021stochastic,burke2021discovering,elkoumy2020directlyfollows}.

Once a process model is enriched with behavioural statistics, it may be a
target
of a reconstruction attack. That is, similar to
reconstruction attacks in machine learning
(ML)~\cite{DBLP:journals/popets/HilprechtHB19,DBLP:journals/ieicet/HidanoMKKH18,DBLP:journals/csur/RigakiG24},
 which strive
for a
characterization of the data used for training the ML model, such an attack
aims at deriving insights about the original process executions. Even if the
exact reconstruction of the executions is not possible, which would yield
severe privacy risks for process stakeholders~\cite{nunez2020quantifying},
it is problematic: The combination of the control-flow of a process model
with behavioural statistics may facilitate conclusions on confidential
information about the underlying business process. For instance, one may
reconstruct dependencies between activity executions and behavioural
patterns, which reveal internal decision procedures that may be exploited
for malicious purposes.

\begin{figure}[t]
\centering
\begin{subtable}[b]{0.68\textwidth}
 \centering
 \scriptsize
\begin{tabular}{l l}
\toprule
ID & Trace\\
\midrule
1 & Register (R), IV Liquid (L), Antibio (A), Rel. D
(D), Return (U)\\
2 & Register  (R), Antibio (A), Antibio (A), Antibio
(A), Rel. E (E)\\
3 & Register  (R), IV Liquid (L), Rel. B (B), Return (U)\\
\bottomrule
\end{tabular}
     \caption{}
         \label{fig:intro_example_log}
     \end{subtable}
\quad
     \begin{subfigure}[b]{.28\textwidth}
         \centering
		\scriptsize
     \scalebox{0.9} {
         	\begin{tikzpicture}
         	\tikzset{level 1/.style={sibling distance=0pt, level
         	distance=30pt}}
         	\Tree[.$\circlearrowleft$
         	[.$\tau$ ]
         	[.{R} ][.{L} ][.{A} ][.{D} ][.{E} ][.{B} ][.{U} ]
         	]
         \end{tikzpicture}}
     \caption{}
         \label{fig:intro_example_flower}
     \end{subfigure}
     \begin{subfigure}[b]{0.42\textwidth}
        \centering
		\scriptsize
		\begin{tikzpicture}
			\tikzset{level 1/.style={sibling distance=0pt, level
			distance=20pt}}
			\tikzset{level 2/.style={sibling distance=0pt, level
			distance=20pt}}
			\Tree[.$\times$
			[.$\rightarrow$ [.{R} ] [.{L} ] [.{A} ] [.{D} ] [.{U} ]]
			[.$\rightarrow$ [.{R} ]  [.{A} ] [.{A} ] [.{A} ] [.{E} ]]
			[.$\rightarrow$ [.{R} ] [.{L} ] [.{B} ] [.{U} ]]
			]
		\end{tikzpicture}
     \caption{}
         \label{fig:intro_example_trace}
     \end{subfigure}
     \begin{subfigure}[b]{0.57\textwidth}
         \centering
     \scriptsize
     \begin{tikzpicture}
     	\tikzset{level 1/.style={sibling distance=0pt, level distance=20pt}}
     	\tikzset{level 2/.style={sibling distance=0pt,level distance=20pt}}
     	\Tree[.$\rightarrow$:3
     	[.{R}:3 ][.$\times$:3 [.$\tau$:1 ][.{L}:2 ]]
     	[.$\circlearrowleft$:3 [.$\tau$:8 ][.{A}:4 ]]
     	[.$\times$:3 [.{D}:1 ][.{E}:1 ][.{B}:1 ]]
     	[.$\times$:3 [.$\tau$:1 ][.{U}:2 ]]
     	]
     \end{tikzpicture}
     \caption{}
         \label{fig:intro_example_middle}
     \end{subfigure}
\vspace{-1.6em}
\caption{
(a) A log of patient treatments and three process models for it:
(b) a `flower model' describing any set of traces;
(c) a `trace model' enumerating all traces; (d) a model offering some
generalization, potentially annotated with frequencies.}
\label{fig:intro_example}
\vspace{-1.8em}
\end{figure}
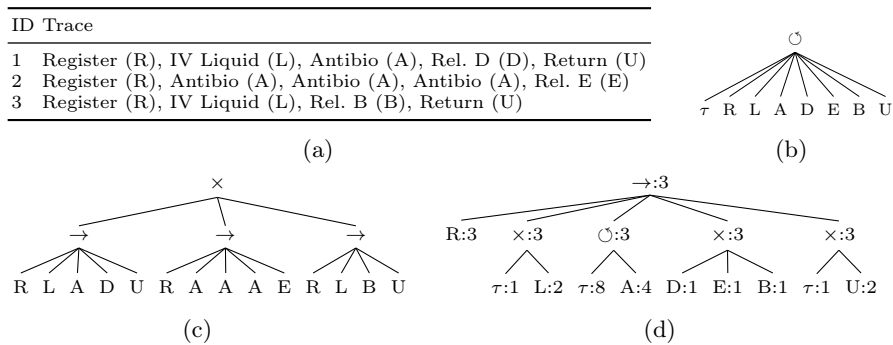

Consider the event log in
\autoref{fig:intro_example_log}, which contains three traces of patient
treatments in a hospital.
The impact of the generalization
adopted in a process model on revealing insights on the original process
executions is illustrated by two
extreme cases: \autoref{fig:intro_example_flower} shows a `flower
model' that represents any log of
traces comprising executions of the respective activities and, hence, does
not enable any conclusions. \autoref{fig:intro_example_trace} shows a `trace
model', which models a lossless representation of each recorded trace variant, but has no information on their probability or frequency. The model represents an infinite number of possible event logs with different frequencies of those traces. Nonetheless, all possible logs include all steps of all process executions for this procedure in the hospital. 
A middle ground is offered by the model in \autoref{fig:intro_example_middle}, which enables
control-flow reconstruction to some extent. In particular, annotating
the model with frequency information reduces number of event logs modeled by this model and reveals certain insights on the
treatments: We conclude that (i) antibiotics have been
given at least twice to a single
patient, (ii) all release types appear to be equally likely, and (iii) there
is at least one patient, who returned after receiving intravenous (IV)
liquid. These insights are shared across the control-flow of all possible logs that this model represents.

In this paper, we study reconstruction attacks on process models and analyse
how the information
contained in process models influences one's ability to reconstruct
the control-flow of process executions. We formulate
reconstruction attacks as play-out strategies for models given as process
trees, which incorporate annotations on branching probabilities or
execution frequencies and especially address the question of how to cope with
repetitive behaviour in the model. Our experiments of applying
the attacks to real-world datasets indicate that
frequency-annotated models of structured processes are particularly
vulnerable.

Below, we first review related work (\autoref{sec:related_work}),
before defining preliminary notions (\autoref{sec:preliminaries}). We then
present our approaches to control-flow reconstruction
(\autoref{sec:reconstruction}), report on our evaluation
(\autoref{sec:evaluation}), and conclude (\autoref{sec:conclusion}).

\section{Related Work}
\label{sec:related_work}

Any attempt to reconstruct the original process executions from a process
model is related to privacy risks, which received much attention in recent
years.
However, we notice that existing work on the quantification of privacy
risks in process mining~\cite{nunez2020quantifying} and the
development of a large number of related privacy-preserving
techniques~\cite{fahrenkrog2023semantics,hildebrant2023pmdg,DBLP:journals/dke/RafieiA21,DBLP:conf/rcis/RafieiWPA23} has
focused primarily on event
logs. As such, there is a reasonable level of understanding of these risks
and possible mitigation strategies.

The risks induced by process models discovered from event logs, in turn,
have been described only recently in~\cite{maatouk2022quantifying}. Here,
the authors
quantify the re-identification risk in frequency annotated
block-structured process models with a two-step approach: First, a
play-out strategy is used to reconstruct event logs from the
process model.
Second, the measures proposed in~\cite{DBLP:conf/icpm/RafieiA20} are used
to
quantify the re-identification risk in the reconstructed log, to then assess the re-identification risk of the original log caused by the release of the process model. However, this
approach
is only feasible if there is a strong similarity between the reconstructed
logs and the original log. This aspect is not further studied
in~\cite{maatouk2022quantifying}, though, which is a research gap that we
close with our work.

Play-out strategies and the comparison of the obtained output with a ground
truth are also studied in other process mining settings:
Conformance checking \cite{DBLP:books/sp/CarmonaDSW18} relates the behaviour
described by a
process model with behaviour in an event log. Yet, often we cannot
fully trust both our model and the source event log, as indicated
in~\cite{DBLP:conf/bpm/Rogge-SoltiSWMG16}. In our context, missing or extra
behaviour in
the process model as assumed in conformance checking would further impair the chance of a successful reconstruction attack.
As such, we assume the process model to be a good
representation of the observed process behaviour, which presents the worst
case for any attempt to derive insights on the underlying business process,
as it simplifies the reconstruction. Both process
simulation~\cite{DBLP:journals/dss/CamargoDG20} and stochastic process
mining~\cite{burke2021stochastic} aim to more accurately capture the underlying
process observed in process executions. These streams of
research investigate how close simulated
process executions~\cite{DBLP:journals/corr/abs-2303-17463} or the probability
distributions in stochastic process model
executions~\cite{DBLP:journals/is/LeemansP23} are to the actual
observations. Unlike our work, however, these approaches do not target the
reconstruction
of the original log, but on representing the
general process behaviour including possible future process executions.

\section{Preliminaries}
\label{sec:preliminaries}
Below, we summarize essential notions for event logs and process trees that are
used in the remainder of the paper.

\mypar{Event Log}
Let $\mathcal{A}$ be the universe of activities. A trace $t \in \mathcal{A}^*$,
where $\mathcal{A}^*$ is the set of all finite sequences over $\mathcal{A}$, is a sequence of
activities. In such a trace, each activity $a$ denotes the recorded event of the execution of a well-defined step in a process.
$\mathcal{T} = \mathcal{A}^*$ denotes the universe of traces.
A trace $t \in  \mathcal{T}$ is represented as $t = \langle a_1, a_2, ..., a_n \rangle$, where $a_1, a_2, ..., a_n \in \mathcal{A}$. With $\abs{t}$ we denote the length of a trace $t \in  \mathcal{T}$, i.e., the number of activities in the trace. Denoting with $\mathcal{B}(X)$ the set of all possible multisets over $X$, let $\mathcal{L} = \mathcal{B}(\mathcal{T})$ be the universe of event logs. An event log $l \in \mathcal{L}$ is a finite multiset of traces.

\mypar{Process Tree}
In this work, we consider process trees as the formal model to capture business
processes.  A process
tree represents a process in a hierarchical (block-structured)
way~\cite{burke2021discovering,DBLP:conf/apn/LeemansFA13}.
Process trees can be transformed into models of other languages for business
processes, such as Petri nets or BPMN models~\cite{vanDerAalst2016process}. As
such, the ideas outlined in the remainder are not limited to process trees.
In general, a process tree denotes a process as a rooted tree. Its leaf nodes
represent
activities and all
other nodes represent operators. Following the aforementioned references, we
formally define a process tree as follows:
\begin{definition}[Process Tree]\label{def:processtree} Let $A \in \mathcal{A}$
be a finite set of activities and let $\tau \not\in A$ denote the silent
activity, which cannot be observed in a trace. A process tree $Q$, is defined
recursively as:
	\begin{itemize}[nosep]
		\item If $a\in A \cup \{\tau\}$, then $Q = a$ is a process tree.
		\item If $n\geq1$, $Q_1, Q_2, \dots, Q_n$ are process trees, and $ \processTreeOp \in \{\to, \times, \wedge\}$, then $Q = \processTreeOp(Q_1, Q_2, \dots, Q_n)$ is a process tree.
		\item If $n\geq2$, $Q_1, Q_2, \dots, Q_n$ are process trees, and $\processTreeOp = \circlearrowleft$, \\then $Q = \processTreeOp (Q_1, Q_2, \dots ,Q_n) $ is a process tree.
	\end{itemize}
\end{definition}
A process tree might be annotated with information about probabilities or
frequencies of the recorded behaviour. We capture such
information by a weight $w\in \mathbb{R}$ that is assigned to a process tree
$Q$, which is denoted by $Q:w$.

\begin{figure}[t]
	\centering
	     \begin{subfigure}[b]{.22\textwidth}
      \centering
	     	\begin{tikzpicture}
	         	\tikzset{level 1/.style={level distance=18pt}}
	         	\tikzset{level 2/.style={level distance=18pt}}
	         	\tikzset{level 3/.style={level distance=18pt}}
	         	\Tree[.$\to$
	         	[.$\wedge$
	         		[.$a$ ][.$\times$
	         			[.$b$ ][.$c$ ]
	         		]
	         	][.$\circlearrowleft$
         			[.$d$ ][.$\tau$ ]
	         	]
	         	]
	         \end{tikzpicture}
	     \end{subfigure}\quad \quad \quad
	\includegraphics[width=0.6\textwidth]{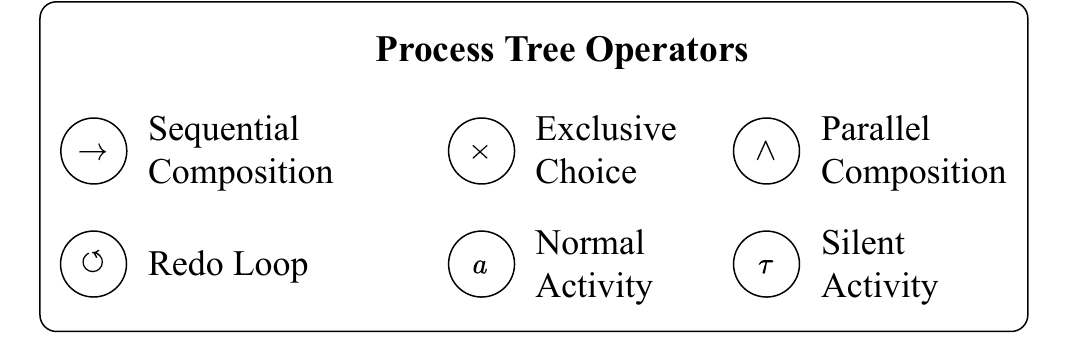}
	\caption{Visualization of the process tree $Q = \ \to ( \wedge ( a,  \times
	(
	b, c)), \circlearrowleft ( d, \tau))$.}
	\label{fig:preliminaries:exampleProcessTree}
    \vspace{-.9em}

\end{figure}

Consider \autoref{fig:preliminaries:exampleProcessTree}, which shows
the process tree $Q = \ \to
( \wedge ( a,  \times ( b, c)), \circlearrowleft ( d, \tau))$. The $\to$
operator refers to the execution of the child nodes in sequential
order, i.e., the execution of
$\wedge ( a,  \times ( b, c))$ is followed by the execution of
$\circlearrowleft ( d, \tau)$.
The $\wedge$ operator defines the execution of all of its child nodes in any
order, while the $\times$ operator specifies an exclusive choice.
The $\circlearrowleft$ operator has at least two children, the first being
the ``do'' part of a loop; all other
children representing ``redo'' parts. The ``do'' part is always executed;
execution of the ``redo'' part is optional and only one of the ``redo''
parts is executed, before the ``do'' part is executed again.

To formalize the semantics of process trees, we need the following auxiliary
operators for general sequences~\cite{vanDerAalst2016process}:

\begin{definition}[Auxiliary Operators]\label{def:sequenceoperators} Let
$\sigma_1, \sigma_2 \in A^*$ be two sequences over $A$ and let $S_1, S_2, \dots
,S_n \subseteq A^*$.  We define two operators as:
	\begin{itemize}[nosep]
		\item Concatenation: $\sigma_1 \cdot \sigma_2 \in A^*$ concatenates two sequences. The concatenation operator can be generalized to sets of sequences by $S_1 \cdot S_2 = \{ \sigma_1 \cdot \sigma_2 \mid \sigma_1\in S_1 \wedge\sigma_2 \in S_2\}$ and $\bigodot_{1 \leq i \leq n} S_i = S_1 \cdot S_2 \dotsm S_n$ concatenates an ordered collection of sets of sequences.
		\item Shuffle:  $\sigma_1 \diamond \sigma_2 \in A^*$ generates the set of all interleaved sequences. The shuffle operator can be generalized to sets of sequences by $S_1 \diamond S_2 = \{ \sigma_1 \diamond \sigma_2 \mid \sigma_1\in S_1 \wedge\sigma_2 \in S_2\}$ and $\lozenge_{1 \leq i \leq n} S_i = S_1 \cdot S_2 \dotsm S_n$ shuffles an ordered collection of sets of sequences.
	\end{itemize}
\end{definition}
Given two sequences $\sigma_1 = \langle a, b \rangle$ and $\sigma_2 = \langle
c, d \rangle$, the operators yield $\sigma_1 \cdot \sigma_2 = \langle
a,b,c,d\rangle$ as well as $\sigma_1 \diamond \sigma_2 =
\{ \langle a,b,c,d \rangle,
\langle a,c,b,d\rangle,
\langle c,a,b,d\rangle,
\langle a,c,d,b\rangle,
\langle c,a,d,b\rangle,$ $
\langle c,d,a,b\rangle  \}$. Furthermore, we define the language of a process tree as follows.

\begin{definition}[Language of a Process Tree]\label{def:languagePT} Let $Q \in \mathcal{Q}$ be a process tree over a set $A$. $\mathcal{L}(Q)$ denotes the language of $Q$, i.e., the set of traces that can be generated. $\mathcal{L}(Q)$ is defined recursively:
	\begin{itemize}[nosep]
		\item  $\mathcal{L}(Q) = \{  \langle a \rangle\}$, if $Q = a \in A$,
		\item  $\mathcal{L}(Q) = \{ \langle \rangle \}$, if $Q = \tau$,
		\item  $\mathcal{L}(Q) = \bigodot_{1 \leq i \leq n} \mathcal{L}(Q_i)$,
		if $Q = \to (Q_1, Q_2, \dots, Q_n)$,
		\item  $\mathcal{L}(Q) = \bigcup_{1 \leq i \leq n} \mathcal{L}(Q_i)$,
		if  $Q = \times (Q_1, Q_2, \dots, Q_n)$,
		\item  $\mathcal{L}(Q) = \lozenge_{1 \leq i \leq n} \mathcal{L}(Q_i)$,
		if $Q = \wedge (Q_1, Q_2, \dots, Q_n)$,
		\item  $\mathcal{L}(Q) =  \{ \sigma_1 \cdot \sigma_1' \cdot  \sigma_2
		\cdot \sigma_2' \dotsm \sigma_m \in A^* \mid m \geq 1 \ \wedge\ \forall
		\ {1
		\leq j \leq m}: \sigma_j \in \mathcal{L}(Q_1) \ \wedge \  \sigma_j' \in
		\bigcup_{2 \leq i \leq n} \mathcal{L}(Q_i)\}$,
		if $Q = \circlearrowleft(Q_1, Q_2, \dots, Q_n)$.
	\end{itemize}
\end{definition}
Based on the definition above, we see that $\mathcal{L}(\circlearrowleft(Q_1,
Q_2, \dots, Q_n)) =
\mathcal{L}(\circlearrowleft(Q_1, \times (Q_2, \dots, Q_n)))$, i.e., a
restriction to a single ``redo'' child does not lower the expressiveness of the
model. As a consequence, without loss of generality, we assume that an
$\circlearrowleft$ operator has only one
``redo'' child in the remainder, to simplify the presentation.
Turning to \autoref{fig:preliminaries:exampleProcessTree}, the language of $Q$
is unbounded due to the loop operator, i.e., $\mathcal{L}(Q) = \{ \langle a,b,d
\rangle, \langle a,c,d \rangle, \langle b,a,d \rangle,  \langle c,a,d
\rangle,   \langle a,b,d, d \rangle, \langle a,c,d,d \rangle\dots\}$.

\section{Control-Flow Reconstruction}
\label{sec:reconstruction}

\begin{figure}[t]
    \centering
    \includegraphics[width=1\textwidth]{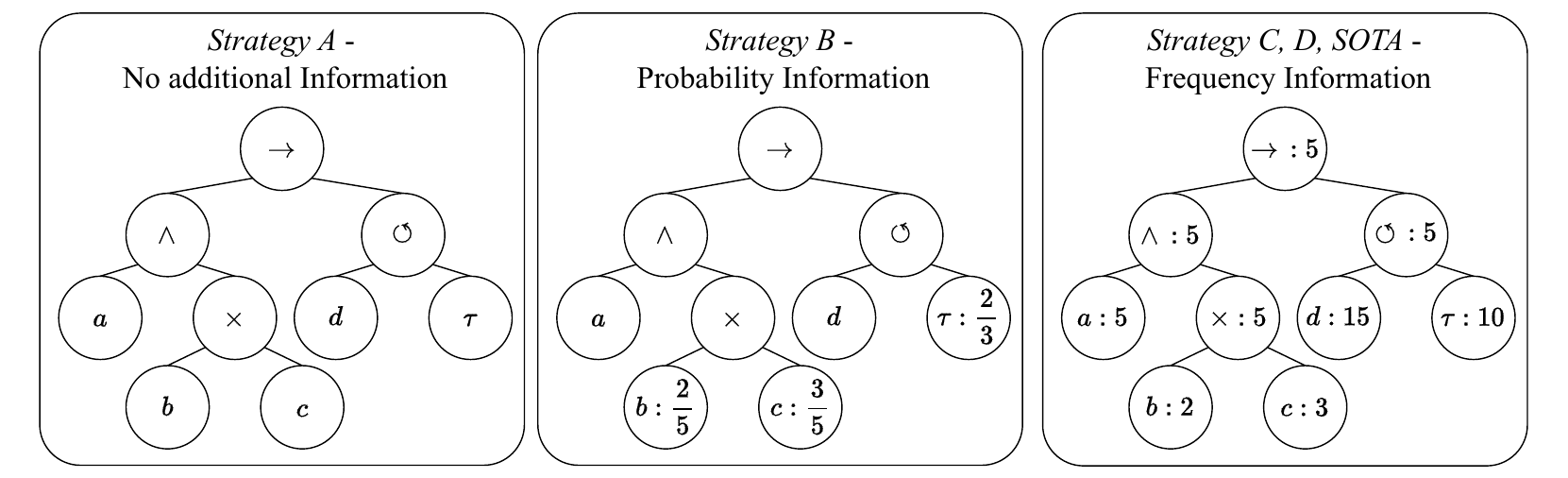}
    \caption{The three common scenarios how a model discovered from log $
L = [\langle a, b, d \rangle,  \langle a, c, d \rangle, \langle c, a, d, d, d, d, d \rangle^2,\langle b, a, d, d, d\rangle]$ can be released.}
    \label{fig:POV_Strategies}
    \vspace{-1.2em}
\end{figure}

As illustrated in our initial example in \autoref{fig:intro_example}, process models may facilitate conclusions on the event log from which the model was discovered. We therefore formulate the respective control-flow reconstruction attacks as five different play-out strategies that, given a process tree, generate a reconstructed event log. We will compare in \autoref{sec:evaluation} the control-flow of the reconstructed log with the control-flow of the original log. The strategies are motivated to reflect the three common ways a process model can be released, see \autoref{fig:POV_Strategies}, and how they utilize this information to reconstruct the control-flow. This enables us to analyze the impact different kinds of additional information, as well as different usage of this information, have on the reconstruction success. %
We first introduce our play-out strategies in
\autoref{sec:strategies_explain}. Next, in \autoref{succ:approxLoop}, we discuss specific issues related to the
handling of loops.

\subsection{Play-out Strategies for Process Trees}
 \label{sec:strategies_explain}

In essence, a play-out strategy defines a
 particular traversal of the process tree according to the control-flow
 structure defined by it.

 \begin{definition}
 Given a process tree $Q$, a play-out strategy $p$ is a function that, applied
 to $Q$, returns an event log $L_p \subseteq \mathcal{B}(2^{\mathcal{L}(Q)})$.
 \end{definition}
Before we formalize the individual aspects of each play-out strategy,
we define some general rules that guide all strategies and apply to any
traversal of a process tree, i.e., the generation of a single trace based on the process tree:

\mycomment{
  \begin{table}[b]
     \centering
     \footnotesize
     \caption{Information required by the presented play-out strategies.}
     \label{tab:required_information}
     \vspace{0.8em}
     \begin{tabular}{l @{\hspace{1cm}} l}
         \toprule
         Strategy & Required Information \\
         \midrule
         Strategy A &  Process tree without annotations \\
         Strategy B & Process tree with relative frequencies  \\
         Strategy C & Process tree with absolute frequencies \\
         Strategy D & Process tree  with absolute frequencies \\
         \midrule
         SOTA strategy~\cite{maatouk2022quantifying} & Process tree with
         absolute frequencies \\
   \bottomrule
     \end{tabular}
 \end{table}

Below, we introduce a total of four play-out strategies for process trees.
The strategies are defined for process trees that include different types of
information on frequencies, as outlined in \autoref{tab:required_information}.
For the sake of completeness, the table also includes the play-out strategy
introduced in~\cite{maatouk2022quantifying}, which we will also describe as an
instantiation of our framework.
}

\begin{itemize}[left=1em]
    \item[$R_0$] Start the traversal with an empty trace.
	\item[$R_1$] If a non-silent leaf node (i.e., not $\tau$) is encountered
	during
	the
	traversal, the respective activity is concatenated to the
	current trace.
	\item[$R_2$] If a silent leaf node ($\tau$) is encountered during the
	traversal,
	the current trace remains unchanged.
	\item[$R_3$] Once the traversal considered all children of a node $Q$, it
	returns
	to and continues with the parent node. If $Q$ is
	the root node, the reconstructed trace will be added to the result.
\end{itemize}
Similarly, we provide some general rules for the play-out of process trees that
relate to the operators for sequential composition and parallel composition. $R_\wedge$ does not apply to the \emph{SOTA Strategy} \cite{maatouk2022quantifying}.
\begin{itemize}[left=1em]
	\item[$R_{\to}$] When $Q=\ \to(Q_1, \dots, Q_n)$
	is
	encountered, the
	traversal continues with the child nodes $Q_1, \dots, Q_n$ in the
	respective order.
	\item[$R_{\wedge}$] \mycomment{ When a node $Q=\wedge(Q_1, \dots, Q_n)$ of a parallel operator
	is
	encountered, the traversal continues with all child nodes $Q_1,
	\dots, Q_n$, creating
	one partial trace $t_1, \dots, t_n$ for each of them. Subsequently, one of
	these traces $t_i$, $1\leq i\leq n$ is
	uniformly chosen at random.  Now these traceses   The procedure is repeated with all remaining
	child nodes $Q_1, \dots, Q_{i-1}, Q_{i+1}, \ldots, Q_n$, until all of them
	have been incorporated. This rule does not apply to the SOTA strategy \cite{maatouk2022quantifying}. }
	When $Q=\wedge(Q_1, \dots, Q_n)$ is encountered, all $U_1, \dots, U_n$ sub-trees are executed until all sub-trees reach a leaf node or $Q$ again, then it is chosen uniformly at random which leaf node is executed. This is repeated until all sub-trees are completely executed and back to $Q$.
    Thus, true parallelization of activities is achieved.
\end{itemize}
Based thereon, we define a first basic play-out strategy that is not based on
any additional information on frequencies.

\mypar{Strategy A} This strategy considers only the semantics of the operators
in a process tree. For the exclusive choice operator and the loop operator, the
respective control-flow choices are taken uniformly at random:
\begin{itemize}[left=1em]
	\item[$R^{A}_{\times}$] When $Q= \times(Q_1, \dots, Q_n)$ is encountered, traversal continues with one child $Q_i$, $1\leq
	i\leq n$, chosen uniformly at random.
	\item[$R^{A}_{\circlearrowleft}$] When $Q= \circlearrowleft(Q_1, Q_2)$ is encountered, traversal continues with the child $Q_1$. Then, a choice
	between executing child
	$Q_2$ and then child $Q_1$ or ending the traversal of
	$Q$ is made. This decision is made with probability $\sfrac{1}{2}$, until the option to end the traversal of $Q$ is taken.

\end{itemize}

\mypar{Strategy B} This strategy interprets the weights assigned to nodes as fixed
branching probabilities. 
These probabilities will be derived from frequencies. For notational purposes, we let the strategy compute these probabilities using the actual frequencies. But the derived fixed probabilities that are computed and used by this strategy correspond to the probabilities a probability-annotated model would have:
\begin{itemize}[left=1em]
\item[$R^{B}_{\times}$] When $Q= \times(Q_1:w_1, \dots, Q_n:w_n):w$ is encountered, traversal continues with one child $Q_i$, $1\leq
i\leq n$, chosen with probability $^{w_i}/_{\sum_{1 \leq j \leq n} w_j}$.
\item[$R^{B}_{\circlearrowleft}$] When $Q= \circlearrowleft(Q_1:w_1, Q_2:w_2):w$
is encountered, we follow the approach from $R^{A}_{\circlearrowleft}$, but adopt the
probability of $1- {}^{w}/_{w_1}$ for the option to continue with children $Q_2$ and
$Q_1$, and $^{w}/_{w_1}$ for the option to end $Q$'s traversal. 
\end{itemize}

\noindent Further strategies leverage the actual frequencies and interpret them in absolute terms.
That is, traversal changes the respective counts, which is captured by the
following rule that applies to all remaining strategies:
\begin{itemize}
	\item[$R_4$] Upon traversal of a node $Q: w$, the value of $w$ will be
	decreased
	by one.
\end{itemize}
Based thereon, we distinguish two strategies to incorporate the absolute
frequencies in the traversal of nodes that model control-flow choices.

\mypar{Strategy C} This strategy takes control-flow choices related to
exclusive choice operators and loop operators, with probabilities that are
determined based on the leftover frequencies:
\begin{itemize}[left=1em]
	\item[$R^{C}_{\times}$] When $Q= \times(Q_1:w_1, \dots, Q_n:w_n):w$ is encountered, traversal continues with one child $Q_i$, $1\leq
	i\leq n$, chosen with probability $^{w_i}/_{\sum_{1 \leq j \leq n} w_j}$.
	Note that this rule differs from the one of \emph{Strategy B}, since the
	weights $w_i$ are continuously updated during traversal, as mentioned above.
	\item[$R^{C}_{\circlearrowleft}$] When $Q= \circlearrowleft(Q_1:w_1, Q_2:w_2):w$
	is encountered, traversal first continues with child $Q_1$. If after this, $w_1 = w_2$ holds, traversal iteratively continues with children $Q_2$ and $Q_1$, until $w_1 = 0$. Intuitively, such an approach collects all leftover frequencies with the last trace that is generated. Otherwise, if $w_1 \neq
	w_2$, we distinguish $w_1 = 0$, in which case traversal of $Q$ ends, and
	$w_1 > 0$, in which case traversal iteratively continues in the loop as in
	\emph{Strategy B}, with probability $1 - {}^{w}/_{w_1}$ for the option including
	the children $Q_2$ and $Q_1$, and ${}^{w}/_{w_1}$ for the option to end
	traversal.
\end{itemize}
\mypar{Strategy D with Variance \emph{v}} This strategy denotes an adaptation
of \emph{Strategy C}. While it also takes all control-flow choices
with probabilities that are determined based on the leftover frequencies,
it includes a normal distribution to decide on the number of loop iterations.
As
usual, by $\mathcal{N}(\mu, \sigma^2)$, we denote a normal distribution with
mean $\mu$ and variance $\sigma^2$. Then, the strategy replaces the rule of
\emph{Strategy C} for the loop operator by the following rule:
\mycomment{
\item[$R^{D}_{L}$] When a node $Q= \circlearrowleft(Q_1:w_1, Q_2:w_2):w$ of
	a loop
	operator
	is encountered, traversal continues with child $Q_1$, if $w_1 = w_2$, as in
	\emph{Strategy C}. Also, if $w_1 \neq
	w_2$ and $w_1 = 0$, again, traversal of $Q$ ends. However,
	if $w_1 \neq  w_2$ and $w_1 > 0$, traversal includes the loop, i.e.,
	children $Q_2$ and $Q_1$ a total of $x$-times, where $x$ is randomly
	sampled from
	the normal distribution, $x \sim \mathcal{N}(\frac{w_1}{w}, v)$.}
\begin{itemize}[left=1em]
 \item[$R^{D}_{\circlearrowleft}$] When $Q= \circlearrowleft(Q_1:w_1, Q_2:w_2):w$ is encountered, traversal continues the same way as in $R^{C}_{\circlearrowleft}$. All rules apply, except for the case
	when $w_1 \neq  w_2$ and $w_1 > 0$. In this case, we will traverse the loop, i.e.,
	children $Q_2$ and $Q_1$ a total of $\min(\lfloor \abs x  \rfloor, w_2) $-times, where $x$ is randomly
	sampled from
	the normal distribution, $x \sim \mathcal{N}(\frac{w_2}{w}, v)$. To get a positive integer, we compute $\lfloor \abs x  \rfloor$. Our experiments have shown that rounding up or concatenating the functions in different order had no measurable impact. We return to the parent node after we traversed the loop $\min(\lfloor \abs x  \rfloor, w_2)$-times or $w_1 = 0$.
\end{itemize}

\mypar{State-of-the-Art (SOTA) Strategy} This strategy was introduced in ~\cite{maatouk2022quantifying}. It traverses a process tree like \emph{Strategy C} but does
a sequential play-out for the parallel composition, hence $R^{SOTA}_{\wedge} = R_{\to}$ , and for the exclusive
choice operator: 
\begin{itemize}[left=2.8em]
	\item[$R^{\mathit{SOTA}}_{\times}$] When $Q= \times(Q_1:w_1, \dots,
	Q_n:w_n):w$ is encountered, consider the child nodes $Q_1, \dots, Q_n$
	in their respective order. Traversal continues with the first child $Q_i$
	with a positive weight, i.e., $w_i>0$.
\end{itemize}

\begin{figure}[t]
    \centering
     \includegraphics[width=1\textwidth]{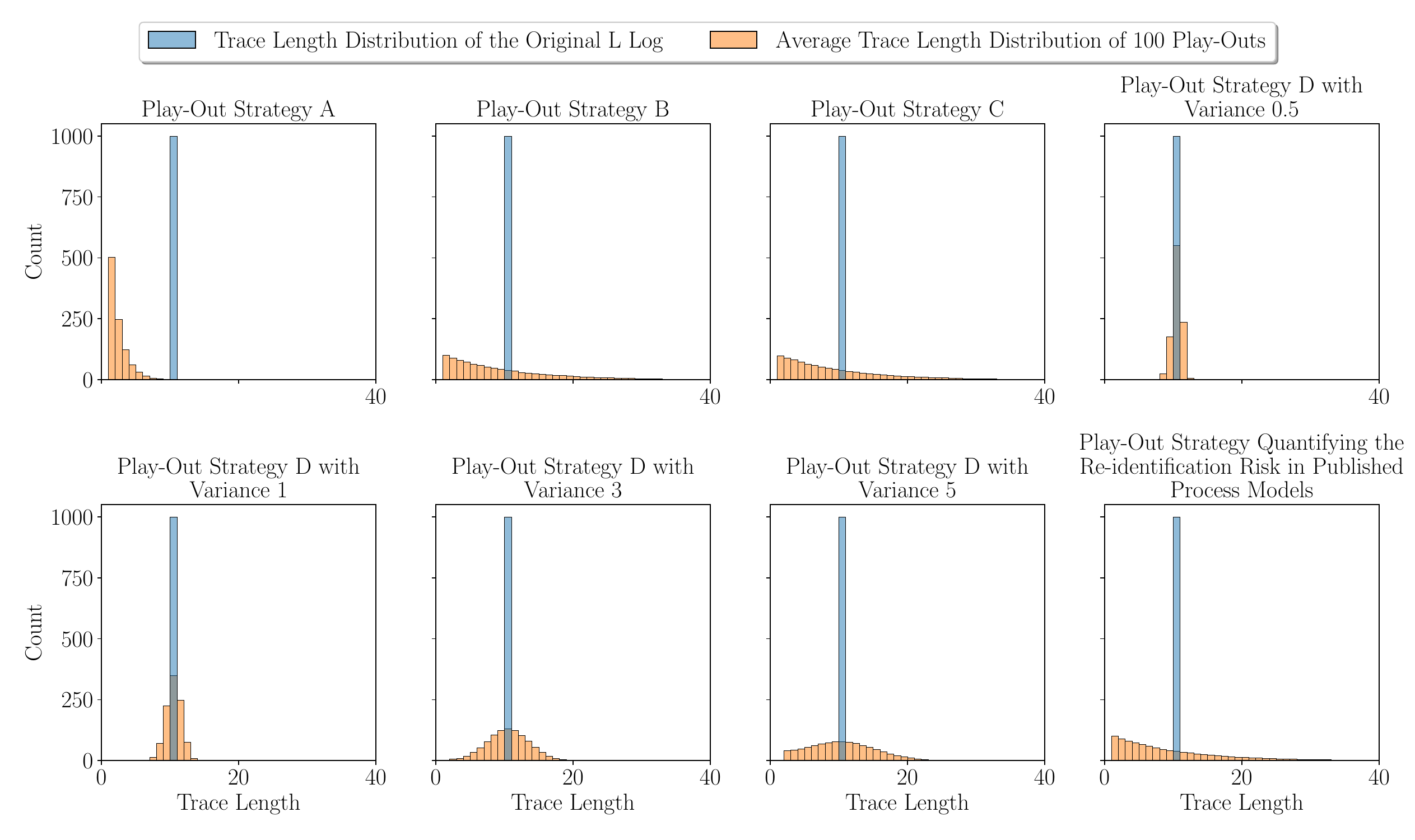}
    \caption{The distribution of the average trace length when playing out
    100 traces using each play-out strategy from $\circlearrowleft (a$:$10000, \tau$:$9000)$:$1000$, along with the distribution of the
    original log $L= [\langle a,a,a,a,a,a,a,a,a,a\rangle^{1000}]$.}
    \label{fig:loopresult}
    \vspace{-1.2em}
\end{figure}

\subsection{Reconstructing the Number of Loop Iterations}
\label{succ:approxLoop}
Next, we discuss the motivation for the approach presented in \emph{Strategy D} that determines the number of loop iterations upfront, instead of relying solely on branching probabilities.

As an illustrative example, consider the process tree $\circlearrowleft (a$:$10000, \tau$:$9000)$:$1000$ discovered from event log $L = [\langle a,a,a,a,a,a,a,a,a,a\rangle^{1000}]$.  \autoref{fig:loopresult} shows the trace length distribution of $L$ and the normalized trace length distribution of 100 play-outs, each containing 1000 traces for each play-out strategy.  The majority of traces produced by all strategies except \emph{Strategy D} are much shorter than the traces of log $L$.
The reason is that these play-out strategies, and modelling techniques such as~\cite{rogge2014discovering} or partly~\cite{burke2021discovering}, capture the execution of the ``redo'' child of a loop operator with some probability $p$. Suppose $p$ is fixed, like in \emph{Strategy A} and \emph{B}. In that case, each iteration is a Bernoulli trial with the number of iterations being a geometric variable \cite{burke2021discovering}. Because in \emph{Strategy C} and the \emph{SOTA Strategy}, the probability changes at each loop iteration, the sequence of iterations is not a sequence of Bernoulli trials. Nonetheless, our experiments show that the resulting distributions of iterations are actually close to a geometric distribution.

To reconstruct traces with consistently more loop iterations, one must decide on the number of loop iterations upfront. When playing out a process tree $Q = \circlearrowleft (Q_1 : w_1, Q_2 : w_2) : w$, we  know that the traces of the original log, took in $w$ $\circlearrowleft$-executions, on average $^{w_2}/_{w}$ loop repetitions. In our example, traces took, on average, $^{9000}/_{1000} = 9$ loop repetitions. \emph{Strategy D} uses this information to set the mean of the normal distribution to $^{w_2}/_{w}$, each time we execute the $\circlearrowleft$ node. Here, the choice of a normal distribution is motivated by the fact that, in each process execution, multiple choices on (re)entering the loop are taken. Once these choices can be assumed to be independent and identically distributed (i.i.d.), the observational error is expected to tend to a normal distribution. %
Even in the absence of knowledge on the variance parameter $v$ of the distribution, we expect it to provide a suitable representation of the number of loop iterations per process execution.

Compared to \emph{Strategy D} the other strategies will perform worse when number of loop iterations is distributed such that the values are large and the variance is low.
\emph{Strategy D} performs worse when the number of loop iterations is distributed with large variance and the values are not centered around the chosen mean.

\section{Experimental Evaluation}
\label{sec:evaluation}

In this section, we evaluate how well our proposed play-out strategies can
reconstruct the control-flow of logs from their discovered models.  We
present our experimental setup in \autoref{sec:exp_setup}, and discuss
evaluation measures in \autoref{secc:measures}. Then, we describe our
results in \autoref{secc:results} and discuss them in
\autoref{secc:discussion}.

\subsection{Experimental Setup}
\label{sec:exp_setup}

\mypar{Experimental Pipeline} We use the inductive miner without noise
filtering to discover the process trees. The lack of noise filtering results in
a perfect fitting process model, a necessary condition to be able to fully reconstruct the
log from the model. In our setting, it is impossible to reconstruct control-flow information about the event log that is not present in the model. To determine the frequency of nodes, we replay each trace of the original event log on the process tree. Each time we visit a node, we increase its weight by one. For
each strategy, we do 100 play-outs of each process tree to obtain the evaluation logs.  For \emph{Strategy A} and \emph{Strategy B}, we fix
the number of traces generated to the number of traces in the original log.
Hence, our results for these strategies are an upper
bound for the reconstruction risks, since usually the number of traces is not
known to the adversary, when the model is not annotated with absolute frequencies.

\mypar{Dataset} %
We evaluate the play-out strategies using four real-world event logs: the BPIC 2015 Municipalities log \cite{vanDongen2015bpi}, the BPIC 2017 log \cite{vanDongen2017bpi}, the BPIC 2013 Closed Problems log \cite{steeman2013bpi}, and the Sepsis log \cite{mannhardt2016sepsis}.
In \autoref{tab:logstatistics} we show certain characteristics of the logs. The logs range from unstructured (BPIC 2015) to structured (BPIC 2013) and also differ drastically in the number of their activities.
In addition to different levels of structuredness, we also considered different trace lengths, since longer traces are potentially harder to reconstruct. The logs differ from having relatively short (BPIC 2013) to very long traces (BPIC 2015).

\mypar{Implementation} Our implementation is available on GitHub\footnote{\url{https://github.com/henrikkirchmann/Control-Flow-Reconstruction}}.
We used the inductive miner and earth mover's distance of
PM4Py~\cite{DBLP:journals/simpa/BertiZS23}.
The runtime of our implemented play-out strategies is fast. On a machine with an AMD Ryzen 5600G a play-out of the BPIC 2013 log is generated in under one second and in 30 seconds one play-out for the BPIC 2017 log.

\begin{table}[t]
	\caption{Descriptive statistics of the event logs.}
	\vspace{0.5em}
	\resizebox{\textwidth}{!}{%
		\begin{tabular}{@{}cccccccc@{}}
			\toprule
			&       &       &        & \multicolumn{3}{c}{\textbf{Trace Length}} &     \\ \cmidrule(lr){5-7}
			\multirow{-2}{*}{\textbf{Event Log}} &
			\multirow{-2}{*}{\textbf{\begin{tabular}[c]{@{}c@{}}\# of \\ Traces\end{tabular}}} &
			\multirow{-2}{*}{\textbf{\begin{tabular}[c]{@{}c@{}}\# of \\ Variants\end{tabular}}} &
			\multirow{-2}{*}{\textbf{\begin{tabular}[c]{@{}c@{}} $\frac{\textbf{\# of Variants}}{\textbf{\# of Traces}}$ \end{tabular}}} &
			\multicolumn{1}{l}{\textbf{Min.}} &
			\multicolumn{1}{l}{\textbf{Avg.}} &
			\multicolumn{1}{l}{\textbf{Max.}} &
			\multirow{-2}{*}{\textbf{\begin{tabular}[c]{@{}c@{}}\# of \\  Activities\end{tabular}}} \\ \midrule
			BPIC 2017                 & 31509 & 15930 & 0.505 & 10          & 38.1          & 180         & 27  \\
			BPIC 2015 Municipalities  & 1199  & 1170  & 0.975 & 2           & 43.5          & 101         & 399 \\
			BPIC 2013 Closed Problems & 1487  & 183   & 0.123 & 1           & 4.4           & 35          & 5   \\
			Sepsis Cases              & 1050  & 846   & 0.805 & 3           & 14.4          & 185         & 17  \\ \bottomrule
		\end{tabular}%
	}
	\label{tab:logstatistics}
	\vspace{-1em}
\end{table}

\subsection{Evaluation measures} \label{secc:measures}

\mypar{Trace Length Distribution}
We look into the trace length distribution to investigate the impact of how different
play-out strategies handle the $\circlearrowleft$ operator. We plot the normalized distributions of each play-out strategy and the distribution of the original log as histograms for the BPIC 2017 log. The plots for the other event logs can be found in our appendix\footnote{\url{https://github.com/henrikkirchmann/Control-Flow-Reconstruction/tree/main/Appendix}}. We calculate the similarity of the distributions using the normalized histogram intersection for all logs:

\begin{definition}[Normalized histogram intersection (NHI)
size]\label{def:NormalizedHistogramIntersection}
	 Let $I$ and $M$ be histograms, each containing $n$ bins, and let $I_j$ respectively $M_j$ denote the number of elements in bin $j$ of $I$ respectively of $M$.  The normalized histogram intersection size is defined to be
		\[ NHI(I,M) = \frac{\sum_{j=1}^{n} \min(I_j,
		M_j)}{\sum_{j=1}^{n}M_j}.\]
\end{definition}

\mypar{Earth Mover's Distance (EMD)}
We compare the reconstructed logs with the original log through the earth mover's distance (EMD) introduced in \cite{leemans2019earth}. The EMD measures the distance, by the Levenshtein distance function, between the trace variant distributions of the event logs. Through the EMD, we can measure the difference between the logs in terms of their control-flow of each trace. This allows us to see if the play-out strategies produce logs that have similar control-flow to the original log, without the need for traces to be the same.

\mypar{Normalized Multiset Intersection (NMI) Size}
The multiset intersection size between two event logs represents the count of
traces from the original log that are completely and successfully
reconstructed. The normalized multiset intersection size, denoted by \(NMI(L_1,
L_2)\), is defined as the sum of the minimum occurrences of each trace
\(\sigma\) in both multi sets \(L_1\) and \(L_2\) divided by $\abs{L_1}$. For example, given the event
logs \(L_1 = [\langle a,b\rangle^2, \langle a,b,c\rangle^2]\) and \(L_2 =
[\langle a,b\rangle^3, \langle a,b,b\rangle]\), we have $NMI(L_1, L_2) = $ $^{2}/_{4}$.
Through this metric, we can determine if the play-out strategies create traces that are exactly the same as the traces of the original log.

\mypar{Reconstructed Eventually Follows Relations}
To check how many dependencies between activities we can reconstruct, we compare the eventually follows relations of the reconstructed logs to the original log.
An eventually follows relation between two activities $a$ and $b$, can be
one of three types: (i) The relation between $a$ and $b$ is of type
\emph{always follows} when in all traces of the log activity $b$ will occur
eventually after $a$; (ii) \emph{sometimes follows} when in some but not all
traces of the log $b$ will occur eventually after $a$;
(iii) \emph{never follows} when in no trace of the log $b$ will  eventually occur after $a$.
To quantify the differences between the predicted eventually follows relations of our play-out strategies and the ones of the original log, we calculate the $F_1$-Scores, as the harmonic mean of precision and recall.

\subsection{Results} \label{secc:results}

\mypar{Trace Length Distribution}
\autoref{tab:nhi_emd_results} shows the NHI size with higher values, meaning better reconstruction of the trace length distribution.
We can observe NHI values above $0.7$ for 3 out of 4 of the event logs, with
the exception being the BPIC 2015 log. Therefore, we can conclude that it is
generally possible to mimic the trace length distribution and to rediscover
general control-flow properties.

\begin{figure}[htbp]
        \vspace{-1.2em}
\centering
        \includegraphics[width=\textwidth]{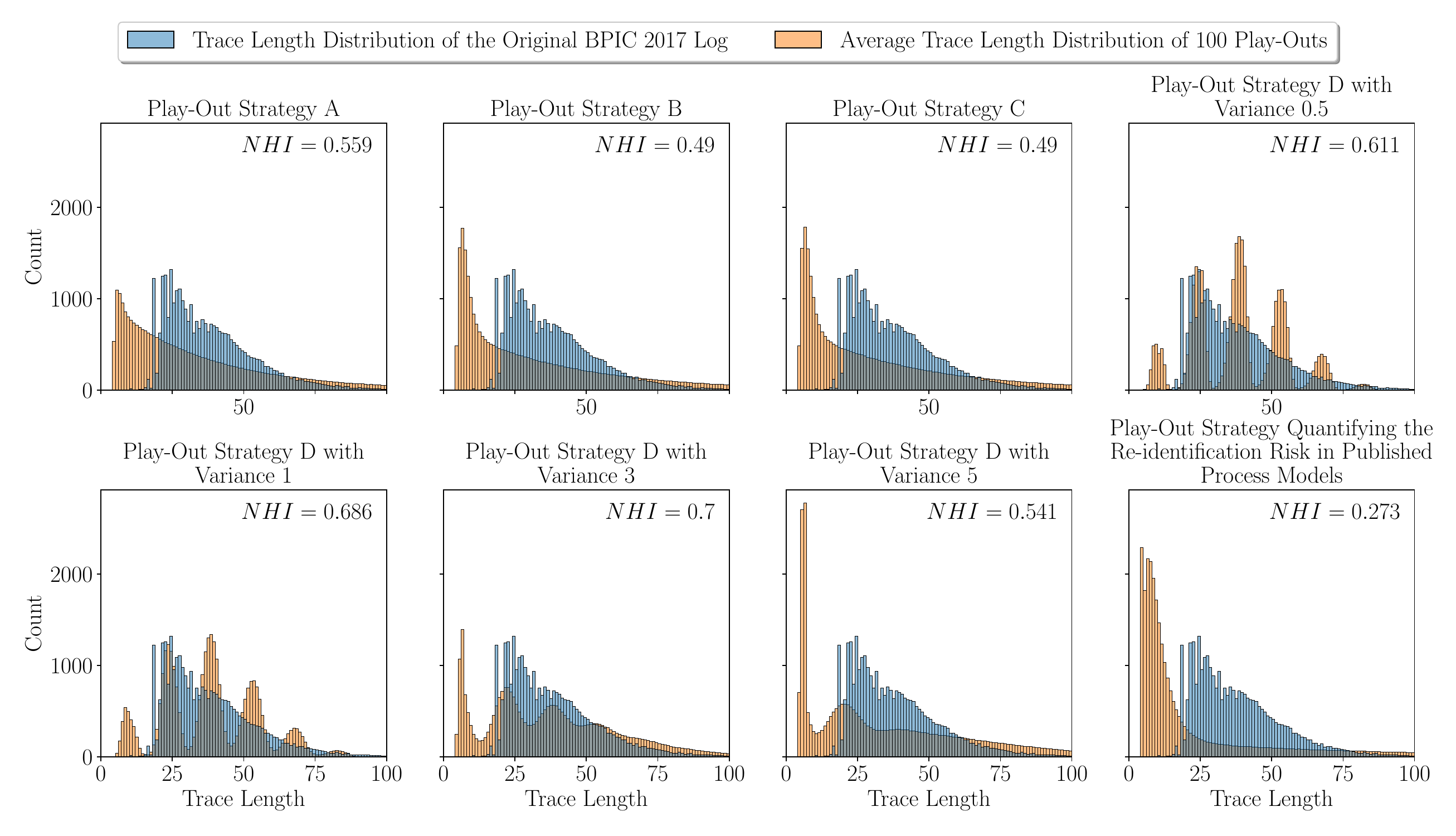}
        \caption{The trace length distributions for the BPIC 2017 log.}
        \label{fig:TraceLengthDistributionBPIC2017}
        \vspace{-1.2em}
\end{figure}

Considering the results in more detail, the success of the reconstruction
might depend highly on the handling of loops. This aspect can be seen by the
difference between the different settings for \emph{Strategy D}. For BPIC
2015 the worst setting (\emph{Strategy D with Variance 0.5}) reached a NHI
of 0.19, while the best setting (\emph{Strategy D with Variance 5}) led to a
NHI value of 0.44.
In \autoref{fig:TraceLengthDistributionBPIC2017}, we can see that, compared to the other strategies, \emph{Strategy D with Variance 0.5-3} creates considerably fewer traces of length below 20.

\begin{table}[t]
	\centering
    \caption{NHI size, EMD and NMI size of 100 play-out logs and the original log. Higher NHI/NMI values and lower EMD values denote higher reconstruction success, the values that indicate the highest reconstruction success are bold.}
	\vspace{0.5em}
    \footnotesize
    \begin{tabular}{l @{\hskip 1.2em} cc @{\hskip 1.2em} ccc @{\hskip 1.2em} cc @{\hskip 1.2em} cc @{\hskip 1.2em} | cc}
    \toprule
    & \multicolumn{2}{c}{BPIC17} & \multicolumn{3}{c}{BPIC13} &  \multicolumn{2}{c}{BPIC15} & \multicolumn{2}{c}{Sepsis} & \multicolumn{2}{|c}{Average} \\
\cmidrule(lr){2-3} \cmidrule(lr){4-6} \cmidrule(lr){7-8} \cmidrule(lr){9-10} \cmidrule(lr){11-12}

        Strategy & NHI & EMD & NHI & EMD & NMI & NHI & EMD & NHI & EMD & NHI & EMD \\
        \midrule
        A & 0.55 & - & 0.66 & 0.35 & 0.19 & 0.17 & 0.93 & 0.50 & 0.74 & 0.47 & 0.67 \\
        B & 0.49 & - & 0.83 & \textbf{0.10} & \textbf{0.59} & 0.43 & 0.87 & \textbf{0.70} & 0.52 & 0.61 & 0.50 \\
        \midrule
        C & 0.49 & - & 0.83 & 0.11 & \textbf{0.59} & 0.43 & 0.87 & \textbf{0.70} & 0.51 & 0.61 & 0.50 \\
        D Var. \sfrac{1}{2} & 0.61 & - & 0.60 & 0.22 & 0.37 & 0.19 & \textbf{0.86} & 0.54 & 0.51 & 0.48 & 0.53 \\
        D Var. 1 & 0.68 & - & 0.66 & 0.18 & 0.44 & 0.22 & \textbf{0.86} & 0.61 & \textbf{0.50} & 0.54 & 0.51 \\
        D Var. 3 & \textbf{0.70} & - & \textbf{0.88} & \textbf{0.10} & \textbf{0.59} & 0.38 & 0.87 & 0.61 & 0.51 & \textbf{0.64} & \textbf{0.49}\\
        D Var. 5 & 0.54 & - & 0.76 & 0.17 & 0.51 & \textbf{0.44} & 0.87 & 0.51 & 0.53 & 0.56 & 0.52\\
        \midrule
        SOTA~\cite{maatouk2022quantifying} & 0.27 & - & 0.83 & 0.14 & 0.52 & 0.38 & 0.92 & 0.69 & 0.62 & 0.54 & 0.56\\
        \midrule
                Avg. for Log & 0.54 & - & 0.75 & 0.17 & 0.47 & 0.33 & 0.88 & 0.60 & 0.55 & 0.55 & 0.53 \\
        \bottomrule
    \end{tabular}
    \label{tab:nhi_emd_results}
    \vspace{-1.5em}
\end{table}

\mypar{Earth Mover's Distance}
\autoref{tab:nhi_emd_results} shows the EMD. Smaller
values, correspond to higher similarity between the control-flow of the play-outs and the original log. Unfortunately, computing the EMD for the BPIC 2017 log was not feasible. We can observe that for the BPIC 2013 log, it
is possible to generate logs that can be very close to the original event log.
However, the BPIC 2015 log shows that this might not be possible for all logs. We can observe that
the difference between the logs is significantly larger than between the
strategies. This lets us conclude that specifics of the process itself
determine the chance of success for the adversary. \emph{Strategy A} that has no additional information about the control-flow performs the worst but is followed by the \emph{SOTA Strategy}, despite having knowledge about the absolute frequencies. The other Strategies reconstruct the control-flow of the original log with similar success in terms of the EMD.

\mypar{Normalised Multiset Intersection Size} For all logs except the BPIC 2013 log, the NMI Size was below 0.01. The values for the BPIC 2013 log are shown in \autoref{tab:nhi_emd_results}. This strongly suggests that the adversary might often not be able to generate the specific traces of the original log. However, for the BPIC 2013 log, we can see that \emph{Strategy A} performs worse than the strategies with knowledge about frequencies.
The results indicate again that knowledge about relative or absolute
frequencies in process models can significantly increase the reconstruction
success. Also, differences between the different settings of \emph{Strategy
D} can be significant.

\mypar{Reconstructed eventually follows relations}
\autoref{tab:ef_f1_results} shows the $F_1$-Scores of the reconstructed
eventually follows relations. Regarding the reconstruction of always follows
(AF) relations, all strategies perform similar, except for \emph{A}, which
performed the worst, and the \emph{SOTA Strategy}, which is the second
worst. Notably, the $F_1$-Scores of \emph{SOTA} are by far the lowest  in 3
out of 4 evaluated logs.
For the sometimes follows (SF) relations, the \emph{SOTA Strategy} again performs the worst, despite having access to absolute frequency information, that is not available to strategies \emph{A} and \emph{B}. Strategy \emph{B} outperforms \emph{D}, despite having only knowledge of branching probabilities and the number of traces to generate.
In the case of never follows (NF) relations, the performance of all
strategies, except for \emph{A}, which performed the worst, is again very
similar.

Overall, we observe a significant level of variance in the
$F_1$-Scores, reaching from cases where no reconstruction is possible to
values as high as 0.89. While it is expected that the highest values are
obtained for the never follows (NF) relations, since they relate to
behaviour that shall not be generated according to the process model, we
also observe relatively high $F_1$-Scores for the always follows (AF)
relations. Those can be interpreted as invariants on the presence of
activity executions, and hence, are particular interesting from a
reconstruction point of view. With $F_1$-Scores around 0.6, we conclude that
a good share of these relations are reconstructed successfully.

\begin{table}[t]
\centering
    \caption{Average $F_1$-scores of 100 play-outs for the reconstructed always (AF), sometimes (SF) and never follows (NF) relations. Higher values denote higher reconstruction success, the highest values are bold.}
	\vspace{0.5em}
    \footnotesize
    \begin{tabular}{l @{\hskip 0.3em} ccc @{\hskip 0.1em} ccc @{\hskip
    0.1em} ccc @{\hskip 0.1em} ccc @{\hskip 0.1em} | ccc}
    \toprule
    & \multicolumn{3}{c}{BPIC17} & \multicolumn{3}{c}{BPIC13} & \multicolumn{3}{c}{BPIC15} & \multicolumn{3}{c}{Sepsis} & \multicolumn{3}{|c}{Average} \\
\cmidrule(lr){2-4}\cmidrule(lr){5-7} \cmidrule(lr){8-10} \cmidrule(lr){11-13} \cmidrule(lr){14-16}
        Strat. & AF & SF & NF & AF & SF & NF & AF & SF & NF & AF & SF & NF &
        AF & SF & NF\\
        \midrule

A & 0.40 & 0.38 & 0.39 & 0.69 & 0.63 & 0.00 & 0.03 & 0.16 & 0.81 & 0.23 & 0.46 & 0.02 & 0.34 & 0.41 & 0.30 \\
        B & 0.52 & 0.54 & 0.47 & \textbf{0.75} & \textbf{0.80} & 0.87 & 0.20 & 0.48 & 0.71 & 0.52 & 0.60 & 0.33 & 0.51 & \textbf{0.61} & 0.60 \\
        \midrule
        C & 0.55 & \textbf{0.55} & 0.47 & \textbf{0.75} & 0.79 & 0.85 & 0.20 & \textbf{0.49} & 0.76 & 0.49 & 0.60 & 0.32 & 0.50 & \textbf{0.61} & 0.60\\
        D \sfrac{1}{2} & \textbf{0.64} & 0.54 & 0.46 & 0.64 & 0.52 & 0.87
        & 0.22 & 0.48 & 0.80 & \textbf{0.59} & \textbf{0.61} & 0.36 &
        \textbf{0.52} & 0.54 & \textbf{0.62}\\
        D 1 & \textbf{0.64} & 0.52 & 0.46 & 0.63 & 0.50 & \textbf{0.89} &
        \textbf{0.23} & 0.47 & 0.79 & 0.57 & 0.60 & 0.35 & \textbf{0.52} &
        0.52 & \textbf{0.62}\\
        D 3 & 0.61 & 0.43 & 0.45 & 0.61 & 0.43 & 0.86 & \textbf{0.23} &
        0.44 & 0.77 & \textbf{0.59} & 0.60 & 0.30 & 0.51 & 0.48 & 0.60\\
        D 5 & 0.62 & 0.43 & 0.46 & 0.61 & 0.43 & 0.86 & \textbf{0.23} &
        0.43 & 0.76 & 0.56 & 0.57 & 0.25 & 0.51 & 0.47 & 0.58\\
        \midrule
        SOTA & 0.16 & 0.30 & \textbf{0.57} & 0.71 & 0.71 & 0.41 & 0.01 & 0.06 & \textbf{0.86} & 0.14 & 0.34 & \textbf{0.53} & 0.44 & 0.35 & 0.59\\
        \midrule
        Avg. & 0.51 & 0.46 & 0.47 & 0.67 & 0.60 & 0.70 & 0.17 & 0.36 & 0.78 & 0.46 & 0.55 & 0.31 & 0.48 & 0.50 & 0.56 \\
        \bottomrule
        \end{tabular}
    \label{tab:ef_f1_results}
    \vspace{-1.3em}
\end{table}

\subsection{Discussion} \label{secc:discussion}

\mypar{Comparison of Play-out Strategies}
Overall, \emph{Strategy A} performed the worst of all play-out strategies.
This is expected, since \emph{Strategy A} lacks information about
probabilities or frequencies in the process model. We conclude that it is
indeed harder or even impossible to successfully reconstruct much of the control-flow of the
original log the un-annotated process model was discovered from.

The play-outs from \emph{Strategy B} and \emph{Strategy C} were almost
similar in our evaluated statistics.
Knowledge of each node's left-over frequency did not help \emph{Strategy C}
to make better reconstruction decisions than \emph{Strategy B}, when
\emph{Strategy B} knows how many traces to generate. This indicates that
when a log with branching probabilities and the number of how many traces
the original log contains are released, the model will reveal nearly the
same amount of control-flow information as it would have done when released
with absolute frequencies.

\emph{Strategy D with Variance $v$} was unable to consistently outperform
\emph{Strategy C} or \emph{Strategy B}.
In our experiments, we could observe that setting the variance value between 1 and 3  led to good results. A limitation of this strategy is that an attacker does not know what variance to pick.
When we sample from the normal distribution with a large variance, like in \emph{Strategy D with Variance $5$} we generate traces that took numerous loop iterations.
The longest trace we generated with \emph{Strategy D with Variance $5$} for the BPIC 2017 log was 863 activities long.
Those long traces consume much of the frequency weights, thus forcing the other reconstructed traces to be shorter.

The \emph{State-of-the-Art Strategy~\cite{maatouk2022quantifying}} performed
worse than \emph{Strategy B} despite knowing the left-over frequencies of
each node. This indicates that we should not execute $\times$ and $\wedge$
nodes sequentially if we want to reconstruct the control-flow from the
original log. We saw for example that this results in many false positive
always follows relations and hence low $F_1$-scores.

\mypar{Assessment of Reconstruction Risk}
In our experiments, we observed that we were able to reconstruct control-flow
properties (trace length distribution). Additionally, we were also able to
generate logs with a small distance to the original log for one process and
a reasonable distance for another. However, we were only able to reconstruct
concrete cases from one log. Finally, we illustrated that information on the
eventually follows relations of the underlying process may be reconstructed
to a significant extent, revealing co-occurrences of activity executions and
their mutual exclusiveness.

However, we acknowledge that, in practice, an attacker also always faces
uncertainty about the reconstructed information, i.e.,
if a reconstructed trace was actually part of the original log. This, in
general, hinders the operationalization of the insights obtained through a
reconstruction attack. This leads to the following assessment
in terms of the reconstruction risk of process models: In general, it is
possible
to retrieve traces from process models. However, this is not possible for
all process models. Therefore, reconstruction risks of process models need
to be considered and taken seriously, but their risk should not be
overstated. Instead, it should be considered that while process models might
not lead to the reconstruction of complete traces, even partially
reconstructed information might be exploitable for an adversary.

\section{Conclusion}
\label{sec:conclusion}
To mitigate confidentially risks, one may resort to publishing a process model instead
of an event log for operational analysis.
In this paper, we argued
that such an approach also potentially incurs risks, since some
information about the original process executions may be reconstructed from the released process model.
We studied this risk and formulated reconstruction
attacks as play-out strategies for models given as process
trees.
We conclude from our experiments that the reconstruction risk
for process trees modelled by the inductive miner from complex real-world event
logs is very low. However, there is a considerable reconstruction risk for more
structured event logs. The annotation of process trees with frequency information increases the reconstruction risk considerably.
Compared to the state of the art, our approaches can consistently provide better results, even with less background knowledge.

In future work, we plan to shift our focus from the quantity of information
that can be reconstructed to a more nuanced analysis. This will involve
examining the specific types of information that can be reconstructed and
the associated uncertainties from an attacker's point of view. Our goal is
to develop algorithms capable of answering questions such as: given a
process model, which traces can be reconstructed that occurred with absolute
certainty in the original log.

\medskip
\noindent
\textbf{Acknowledgements.} This work was supported by the German Federal Ministry of Education and Research (BMBF), grant number 16DII133 (Weizenbaum-Institute).

\bibliographystyle{splncs04}
\bibliography{bibliography}

\end{document}